\documentclass[conference]{IEEEtran}
\IEEEoverridecommandlockouts
\usepackage{cite}
\usepackage{dblfloatfix}
\usepackage{subcaption}
\usepackage{overpic}
\usepackage{amsmath,amssymb,amsfonts}
\usepackage{graphicx}
\usepackage{textcomp}
\usepackage{xcolor}
\usepackage{float}
\usepackage{amsthm}
\usepackage{graphicx}
\usepackage{epstopdf}
\usepackage{amsmath,bm}
\usepackage{amsfonts}
\usepackage{amssymb}
\usepackage{color}
\usepackage{multirow}
\usepackage{multicol}
\usepackage{soul,xcolor}
\usepackage{algorithm}
\usepackage{algpseudocode}

\usepackage{comment}

\theoremstyle{plain}

\newcommand{\vect}[1]{\mathbf{#1}}

\def\Htran{\mbox{\tiny $\mathrm{H}$}}
\def\Ttran{\mbox{\tiny $\mathrm{T}$}}
\def\CN{\mathcal{N}_{\mathbb{C}}} 
\def\imagunit{\mathsf{j}}

\def\imaginary{\mathsf{j}} 
\def\imagunit{\imaginary} 

\begin{document}

\title{GNN-Based Joint Split Processing and Fronthaul Bit Allocation for Cell-Free Massive MIMO}

\author{\"Ozlem Tu\u{g}fe Demir and Parisa Ramezani  
 
\thanks{  \"O. T. Demir is with the Department of Electrical and Electronics Engineering, Bilkent University, Ankara, T\"urkiye (ozlemtugfedemir@bilkent.edu.tr). P. Ramezani is with the School of Electrical, Mechanical and Biomedical Engineering, University of Technology Sydney, Sydney, Australia (parisa.ramezani@uts.edu.au). \newline \indent
The work by \"O. T. Demir was supported by The Scientific and Technological Research Council of Türkiye (TUBITAK) BIDEB
2232-B International Fellowship for Early Stage Researchers under Grant Number 122C149. The work by \"O. T. Demir was supported by the BAGEP Award of the
Science Academy //with funding supplied by Özgül ve Armağan Çağlayan Vakfı.  
}
}

\maketitle
\begin{abstract}
This paper considers a cell-free massive MIMO (multiple-input multiple-output) network whose access points (APs) are connected to a central processing unit (CPU) through capacity-limited fronthaul links. We jointly design the uplink split combining, the downlink split precoding, and the per-beam fronthaul bit allocation to maximize the weighted energy efficiency (WEE) under a total bit budget, by using an end-to-end power model that accounts for the fronthaul transport, the baseband processing at the APs and the CPU, the sleep mode of the unused APs, and the transmit powers. We solve the resulting combinatorial problem with a permutation-equivariant bipartite graph neural network (GNN), which maps the large-scale channel statistics to the bit vector and is trained without labels. A softmax head with a null slot satisfies the budget by construction and allows the network to leave part of it unused. Numerical results show that the learned policy improves the WEE by up to $3.4\times$ compared to a uniform bit allocation. 
\end{abstract}


\vspace{-2mm}

\section{Introduction}

Cell-free massive MIMO (multiple-input multiple-output) replaces the cell-centric architecture with a large
number of geographically distributed access points (APs) that jointly serve all user equipments (UEs) under the
coordination of a central processing unit (CPU), thereby providing almost uniformly good service and macro
diversity \cite{cell-free-book}. However, in practice, every AP exchanges its signals with the CPU over a fronthaul link of finite capacity and, hence, the content and the resolution of the fronthaul signals must be carefully designed \cite{khorsandmanesh2023optimized,ramezani2024joint}. If an AP with $N$ antennas forwards all its $N$ streams at full resolution, the fronthaul rate grows proportionally to $N$ and the CPU has to process the full $LN$-dimensional signal, where $L$ is the number of APs. On the other hand, the channel gains of the individual spatial directions are highly heterogeneous and most of them carry almost no energy.

Split processing exploits this heterogeneity. Recently, splitting architectures have been proposed for multi-user MIMO in \cite{yasamanSplit}, where the radio unit projects the signal onto a lower-dimensional subspace, and only the resulting streams are quantized and transported. Since the projection is designed from long-term channel
statistics, it varies slowly and does not itself burden the fronthaul. In particular, fewer transported bits reduce the fronthaul power, while fewer transported streams reduce the baseband processing power since both the AP-side transform and the CPU-side combining and precoding act on the reduced dimension. In a cell-free network, these savings scale with $L$. Furthermore, an AP whose streams are all dropped can be put to sleep. 

To the best of the authors' knowledge, this paper is the first work that applies split processing to cell-free massive MIMO and learns the resulting design variables. The main contributions are as follows. (i) We formulate the uplink split combining, the downlink split precoding, and the per-beam fronthaul bit allocation as a single discrete decision. In a fixed discrete Fourier transform (DFT) beam basis, the split matrices are beam-selection matrices and a beam that receives zero bits is not forwarded. Hence, the beam selection is induced by the bit allocation. (ii) We model the energy efficiency (EE) with an end-to-end power model that accounts for the fronthaul transport, the baseband processing at the APs and the CPU, the sleep mode of the unused APs, and the transmit powers. (iii) We learn the mapping from the large-scale channel statistics to the bit vector with a permutation-equivariant bipartite graph neural network (GNN), which is trained without labels to maximize the weighted EE (WEE). A softmax head with a null slot satisfies the fronthaul budget by construction and allows the network to spend only part of it, which is the energy-optimal behavior beyond a moderate budget. 

\vspace{-1mm}
\section{System Model}\label{sec:model}
\vspace{-1mm}
We consider a cell-free massive MIMO system in which $L$ APs, each with a
uniform linear array (ULA) of $N$ antennas and a fully digital transceiver,
serve $K$ single-antenna UEs through capacity-limited fronthaul links to a
CPU. We aim to jointly optimize the uplink split-combining structure, the downlink split-precoding structure, and the quantization resolution of every forwarded stream.

The channel between UE $k$ and AP $l$ is
$\mathbf{h}_{kl}\sim\mathcal{CN}(\mathbf{0},\mathbf{R}_{kl})$ with
$\mathbf{R}_{kl}=\beta_{kl}\widetilde{\mathbf{R}}_{kl}$,
where $\beta_{kl}$ is the large-scale fading coefficient and
$\widetilde{\mathbf{R}}_{kl}$ is the normalized correlation matrix;
$\mathbf{h}_{k}=[\mathbf{h}_{k1}^{\Ttran}\cdots\mathbf{h}_{kL}^{\Ttran}]^{\Ttran}$
is the collective channel. The CPU knows the channels perfectly.\footnote{This paper isolates the
interaction between split processing and fronthaul bit allocation from channel
estimation errors and the associated pilot overhead, which are left for future work.}

\subsection{Beam Domain and the Split as Beam Selection}\label{sec:beam}

Each AP works in the fixed unitary DFT (angular beamspace) basis
$[\mathbf{U}_{N}]_{n,m}=\frac{1}{\sqrt{N}}e^{-\imagunit 2\pi nm/N}$ \cite{yasamanSplit}, and
$\mathbf{U}=\mathrm{diag}(\mathbf{U}_{N},\dots,\mathbf{U}_{N}) \in \mathbb{C}^{LN \times LN}$. The
beam-domain channels are
\begin{equation}\label{eq:beamchan}
\mathbf{c}_{kl}\triangleq\mathbf{U}_{N}^{\Htran}\mathbf{h}_{kl},\qquad
\mathbf{c}_{k}=\mathbf{U}^{\Htran}\mathbf{h}_{k}\in\mathbb{C}^{LN},
\end{equation}
and, since the array is a ULA, $\mathbf{c}_{kl}$ is approximately sparse, i.e., a UE illuminates only the few DFT bins aligned with its angles. The split exploits this sparsity in the beam domain.

Since the capacity-limited fronthaul cannot support all $N$ beams of every AP at full resolution, each AP forwards/receives only the beams that carry energy. Hence, both the fronthaul rate and the dimension of the combining/precoding problem at the CPU scale with the number of retained beams rather than with $LN$.

The uplink split-combining matrix $\mathbf{B}_{l}^{\mathrm{UL}}$ and the
downlink split-precoding matrix $\mathbf{B}_{l}^{\mathrm{DL}}$ are restricted
to be \emph{beam-selection} matrices, built from the rows (resp. columns) of
$\mathbf{I}_{N}$ indexed by the forwarded beam sets
$\mathcal{S}_{l}^{\mathrm{UL}},\mathcal{S}_{l}^{\mathrm{DL}}\subseteq
\{1,\dots,N\}$, of cardinality $M_{l}^{\mathrm{UL}},M_{l}^{\mathrm{DL}}$.
Since $\mathbf{U}_{N}$ is unitary, the system is described entirely in the
beam domain: an AP forwards a \emph{subset} of its $N$ beams and quantizes
each of them individually. Let
$b_{l,n}^{\mathrm{UL}},b_{l,n}^{\mathrm{DL}}\in\mathcal{B}\triangleq
\{0,\dots,b_{\max}\}$ be the assigned resolutions. A stream with \emph{zero}
bits carries no information and is not forwarded, so we \emph{define}
\begin{equation}\label{eq:selection}
\mathcal{S}_{l}^{\mathrm{UL}}=\left\{n: b_{l,n}^{\mathrm{UL}}\ge 1\right\},\qquad
\mathcal{S}_{l}^{\mathrm{DL}}=\left\{n: b_{l,n}^{\mathrm{DL}}\ge 1\right\}.
\end{equation}
Hence, the split is induced by the bit vector $\mathbf{b}\in\mathcal{B}^{2LN}$, and the split design and the bit allocation become a single discrete decision. We index the $2LN$ beam slots by $j$.

\subsection{Quantization Model}\label{sec:aqnm}

Every forwarded sample is quantized by a Lloyd--Max quantizer \cite[Tab.~4.1]{jain1989fundamentals}. With the
additive quantization noise model (AQNM) \cite{Fletcher2007a}, an input $z$ of variance
$c_{z}=\mathbb{E}\{|z|^{2}\}$ becomes
\begin{align}\label{eq:aqnm}
&\widetilde{z}=\alpha(b)\,z+q,\qquad \alpha(b)=1-\rho(b),\\
&\mathbb{E}\{|q|^{2}\}=\rho(b)\alpha(b)\,c_{z},
\end{align}
with $q$ uncorrelated with $z$. We use the \emph{tabulated} Lloyd--Max
distortion factors, which are far more accurate than the high-resolution
formula \cite{quantization} at the low resolutions of interest here,
\begin{equation}\label{eq:rho}
\rho(b)=\begin{cases}
\rho_{\mathrm{tab}}[b], & 0\le b\le5,\\[2pt]
\kappa_{q}2^{-2b},\ \kappa_{q}=\tfrac{\pi\sqrt{3}}{2}, & b\ge6,
\end{cases}
\end{equation}
where $\rho_{\mathrm{tab}}=[1,\,0.3634,\,0.1175,\,0.0345,\,0.0095,\,0.0025]$
for $b=0,\dots,5$ \cite[Tab.~4.1]{jain1989fundamentals}. Note that $\rho(0)=1$, i.e., $\alpha(0)=0$, and, hence, a zero-bit beam is switched off by the AQNM gain itself. Therefore, \eqref{eq:selection} holds automatically and no separate selection variable is needed.

\subsection{Uplink Split Combining}\label{sec:ul}

AP $l$ receives $\mathbf{y}_{l}=\sum_{k}\sqrt{p}\,\mathbf{h}_{kl}s_{k}
+\mathbf{n}_{l}$ with symbol $s_{k}\sim\mathcal{CN}(0,1)$, the uplink power $p$, and $\mathbf{n}_{l}\sim \CN(\vect{0}, \sigma^2\vect{I}_N)$, transforms it into the beam domain, applies the split, and quantizes the retained entries as
\vspace{-1mm}
\begin{equation}\label{eq:ulchain}
\mathbf{r}_{l}=\mathbf{U}_{N}^{\Htran}\mathbf{y}_{l},\quad
\mathbf{z}_{l}^{\mathrm{UL}}=\mathbf{B}_{l}^{\mathrm{UL}}\mathbf{r}_{l},\quad
\widetilde{\mathbf{z}}_{l}^{\mathrm{UL}}=\mathbf{A}_{l}^{\mathrm{UL}}\mathbf{z}_{l}^{\mathrm{UL}}
+\mathbf{q}_{l}^{\mathrm{UL}}, \vspace{-1mm}
\end{equation}
with $\mathbf{A}_{l}^{\mathrm{UL}}=\mathrm{diag}(\alpha(b_{l,n}^{\mathrm{UL}}):
n\in\mathcal{S}_{l}^{\mathrm{UL}})$, and forwards the $M_{l}^{\mathrm{UL}}$
quantized streams to the CPU. Collecting the $L$ APs and re-embedding the
dropped beams as zeros (which is exact, since $\alpha(0)=0$), 
with $\alpha_{j}^{\mathrm{UL}}=\alpha(b_{j}^{\mathrm{UL}})$ and
$\mathbf{A}^{\mathrm{u}}=\mathrm{diag}(\alpha_{1}^{\mathrm{UL}},\dots,\alpha_{LN}^{\mathrm{UL}})$, the CPU observes
\vspace{-1mm}
\begin{equation}\label{eq:ulcpu}
\widetilde{\mathbf{z}}^{\mathrm{UL}}=\mathbf{A}^{\mathrm{u}}\left(\textstyle\sum_{k=1}^K\sqrt{p}\,
\mathbf{c}_{k}s_{k}+\mathbf{n}\right)+\mathbf{q}^{\mathrm{UL}}, \vspace{-1mm}
\end{equation}
where the zero-bit slots vanish because $\alpha_{j}^{\mathrm{UL}}=0$. Conditioned on the
channels, the per-slot receive power is
\vspace{-2mm}
\begin{equation}\label{eq:czul}
c_{z,j}^{\mathrm{UL}}=p\sum_{k=1}^{K}\big|[\mathbf{c}_{k}]_{j}\big|^{2}+\sigma^{2},
\vspace{-1mm}
\end{equation}
so the quantization noise plus the amplified receiver noise has the diagonal
covariance\footnote{The AQNM treats the quantization noise as 
independent across and with the quantizer inputs. Under an exact Bussgang
decomposition this is not true as the distortion is only uncorrelated with the
input, and it remains correlated across
beams. We adopt the AQNM only for the
\emph{design} (the combiner, the precoder and the training objective, which
must be differentiable and cheap to evaluate in every step); the reported
performance is instead obtained from an exact Monte-Carlo Bussgang evaluation
with the actual Lloyd--Max quantizers, so the model mismatch is fully
accounted for in Section~\ref{sec:results}.}
\vspace{-1mm}
\begin{equation}\label{eq:Dul}
\mathbf{D}=\mathrm{diag}\left(\alpha_{j}^{\mathrm{UL}}\left(1-\alpha_{j}^{\mathrm{UL}}\right)c_{z,j}^{\mathrm{UL}}
+\left(\alpha_{j}^{\mathrm{UL}}\right)^{2}\sigma^{2}\right)_{j=1}^{LN} .
\vspace{-1mm}
\end{equation}
With the effective channel $\mathbf{g}_{k}=\mathbf{A}^{\mathrm{u}}\mathbf{c}_{k}$ seen by
the CPU, the signal-to-interference-plus-noise ratio (SINR)-optimal minimum mean-squared error (MMSE) combiner is
\vspace{-1mm}
\begin{equation}\label{eq:mmse}
\mathbf{v}_{k}=\left(\sum_{i=1}^{K}p\,\mathbf{g}_{i}\mathbf{g}_{i}^{\Htran}
+\mathbf{D}\right)^{-1}\sqrt{p}\,\mathbf{g}_{k}, \vspace{-1mm}
\end{equation}
giving
\vspace{-1mm}
\begin{equation}\label{eq:seul}
\gamma_{k}^{\mathrm{UL}}=p\,\mathbf{g}_{k}^{\Htran}\left(\sum_{i\ne k}
p\,\mathbf{g}_{i}\mathbf{g}_{i}^{\Htran}+\mathbf{D}\right)^{-1}\!\!\!\!\!\! \mathbf{g}_{k}, \, \,
 \mathrm{SE}_{k}^{\mathrm{UL}}=\log_{2}(1+\gamma_{k}^{\mathrm{UL}}).
 \vspace{-1mm}
\end{equation}
By the matrix inversion lemma, \eqref{eq:seul} is evaluated through the
$K\times K$ matrix $\mathbf{M}=p\,\mathbf{G}^{\Htran}\mathbf{D}^{-1}
\mathbf{G}$, $\mathbf{G}=[\mathbf{g}_{1},\dots,\mathbf{g}_{K}]$, as
$\mathrm{SE}_{k}^{\mathrm{UL}}=-\log_{2}(1-u_{k})$ with
$u_{k}=[(\mathbf{I}_{K}+\mathbf{M})^{-1}\mathbf{M}]_{kk}$, which makes the objective computationally cheap to differentiate in every training step.

\vspace{-2mm}
\subsection{Downlink Split Precoding}\label{sec:dl}

Define $\mathbf{p}_k \in \mathbb{C}^{LN}$ as the power-scaled beam-domain precoding vector for UE $k$. The CPU forms, on every selected beam, the precoded sample
$[\mathbf{z}^{\mathrm{DL}}]_{j}=\sum_{k}[\mathbf{p}_{k}]_{j}s_{k}$, quantizes it with
$b_{j}^{\mathrm{DL}}$ bits and sends it to AP $l$, which expands it back to the
antenna domain and transmits
\begin{equation}\label{eq:dlchain}
\mathbf{x}_{l}=\mathbf{U}_{N}\mathbf{B}_{l}^{\mathrm{DL}}
\widetilde{\mathbf{z}}_{l}^{\mathrm{DL}},\qquad
\widetilde{\mathbf{z}}_{l}^{\mathrm{DL}}=\mathbf{A}_{l}^{\mathrm{DL}}\mathbf{z}_{l}^{\mathrm{DL}}
+\mathbf{q}_{l}^{\mathrm{DL}},
\end{equation}
where $\mathbf{z}_{l}^{\mathrm{DL}}$ collects the entries of $\mathbf{z}^{\mathrm{DL}} $ associated with the selected downlink beams of AP $l$, and  $\mathbf{A}_{l}^{\mathrm{DL}}=\mathrm{diag}(\alpha(b_{l,n}^{\mathrm{DL}}):
n\in\mathcal{S}_{l}^{\mathrm{DL}})$. With
$\mathbf{C}=[\mathbf{c}_{1},\dots,\mathbf{c}_{K}]^{\Htran}$ restricted to the
selected beams, we use the regularized zero-forcing (RZF) precoder
\vspace{-1mm}
\begin{equation}\label{eq:rzf}
\overline{\mathbf{P}}=\mathbf{C}^{\Htran}\big(\mathbf{C}\mathbf{C}^{\Htran}
+\lambda\mathbf{I}_{K}\big)^{-1}\in\mathbb{C}^{LN\times K},
\vspace{-1mm}
\end{equation}
whose rows vanish on the unselected beams, and whose radiated power at AP $l$
is $P_{l}=\sum_{k}\sum_{n\in\mathcal{S}_{l}^{\mathrm{DL}}}
|[\overline{\mathbf{p}}_{k}]_{(l-1)N+n}|^{2}$. To preserve the average per-AP power budget, a \emph{single common} scaling is applied,
\vspace{-1mm}
\begin{equation}\label{eq:scale}
\mathbf{p}_{k}=\xi\,\overline{\mathbf{p}}_{k},\qquad
\xi=\sqrt{P_{\max}\big/\max_{l}\mathbb{E}\{P_{l}\}}, \vspace{-1mm}
\end{equation}
 so the most loaded AP operates
exactly at $P_{\max}$ while the RZF interference geometry is preserved. With
$\alpha_{j}^{\mathrm{DL}}=\alpha(b_{j}^{\mathrm{DL}})$,
$\mathbf{A}^{\mathrm{d}}=\mathrm{diag}(\alpha_{j}^{\mathrm{DL}})$ and
$c_{z,j}^{\mathrm{DL}}=\sum_{k}|[\mathbf{p}_{k}]_{j}|^{2}$ the quantizer input
variance, UE $k$ attains
\vspace{-1mm}
\begin{equation}\label{eq:sedl}
\gamma_{k}^{\mathrm{DL}}=
\frac{\big|\mathbf{c}_{k}^{\Htran}\mathbf{A}^{\mathrm{d}}
\mathbf{p}_{k}\big|^{2}}
{\displaystyle\sum_{i\ne k}\big|\mathbf{c}_{k}^{\Htran}
\mathbf{A}^{\mathrm{d}}\mathbf{p}_{i}\big|^{2}
+\sum_{j}|[\mathbf{c}_{k}]_{j}|^{2}\alpha_{j}^{\mathrm{DL}}
(1-\alpha_{j}^{\mathrm{DL}})c_{z,j}^{\mathrm{DL}}+\sigma^{2}},
\vspace{-1mm}
\end{equation}
and $\mathrm{SE}_{k}^{\mathrm{DL}}=\log_{2}(1+\gamma_{k}^{\mathrm{DL}})$. The middle term in the denominator is the quantization distortion.

\subsection{Power Consumption and Weighted EE}\label{sec:power}

Each active stream is sampled at $f_{s}$ complex samples/s and its in-phase
and quadrature components are each quantized with $b_{j}$ bits, so the
aggregate fronthaul rate and the corresponding power are
\vspace{-1mm}
\begin{equation}\label{eq:pfh}
R_{\mathrm{FH}}=2f_{s}\sum_{j=1}^{2LN}b_{j},\qquad
P_{\mathrm{FH}}=\xi_{\mathrm{FH}}R_{\mathrm{FH}}, \vspace{-1mm}
\end{equation}where $\xi_{\mathrm{FH}}$ [W/(bit/s)] is the fronthaul power-efficiency
coefficient, i.e., the power consumed by the fronthaul transceivers per unit
of transported rate. The CPU processes only what it receives. With the numbers of active streams
\vspace{-1mm}
\begin{equation}\label{eq:streams}
M^{\mathrm{UL}}=\big|\{j:b_{j}^{\mathrm{UL}}\ge1\}\big|,\qquad
M^{\mathrm{DL}}=\big|\{j:b_{j}^{\mathrm{DL}}\ge1\}\big|,
\vspace{-1mm}
\end{equation}
the combining and precoding at the CPU act on vectors of these \emph{reduced}
dimensions rather than on the full $LN$-dimensional ones: dropping a beam
removes one row of $\mathbf{G}$ and one row of $\overline{\mathbf{P}}$, i.e.,
$K$ complex multiply-accumulates per sample. The same holds at the APs, which
evaluate only the $M_{l}^{\mathrm{UL}}$ rows of
$\mathbf{U}_{N}^{\Htran}\mathbf{y}_{l}$ that they forward and, in the
downlink, only the $M_{l}^{\mathrm{DL}}$ columns of $\mathbf{U}_{N}$ selected
by $\mathbf{B}_{l}^{\mathrm{DL}}$ in \eqref{eq:dlchain}, i.e., $N$ complex
multiply-accumulates per sample and per retained beam. Both the beam-domain
transform at the AP and the combining/precoding at the CPU are therefore
proportional to the number of \emph{retained} beams and vanish for a dropped
one, and we lump their per-stream cost into a single coefficient
$\xi_{\mathrm{BB}}$ [W], so that
\vspace{-1mm}
\begin{equation}\label{eq:pbb}
P_{\mathrm{BB}}=\xi_{\mathrm{BB}}\big(M^{\mathrm{UL}}+M^{\mathrm{DL}}\big)
\vspace{-1mm}
\end{equation}
scales with the retained beams rather than all $LN$ beams. Hence, $P_{\mathrm{FH}}$ decreases with fewer bits and $P_{\mathrm{BB}}$ decreases with fewer beams. An AP that forwards no beam
at all, in either direction, is put to sleep and then draws only a fraction
$\eta_{\mathrm{sl}}\in[0,1)$ of its fixed power: with
\begin{equation}\label{eq:awake}
A=\big|\{l:\exists n,\ b_{l,n}^{\mathrm{UL}}\ge1\ \text{or}\
b_{l,n}^{\mathrm{DL}}\ge1\}\big|
\end{equation}
awake APs, the load-independent power is
\begin{equation}\label{eq:pstat}
P_{\mathrm{static}}=P_{\mathrm{CPU}}+P_{\mathrm{AP}}
\big(A+\eta_{\mathrm{sl}}(L-A)\big).
\end{equation}
 Finally, with
$\eta_{\mathrm{UE}},\eta_{\mathrm{AP}}$ the power-amplifier efficiencies,
\begin{equation}\label{eq:ptx}
P_{\mathrm{TX}}=\frac{Kp}{\eta_{\mathrm{UE}}}
+\frac{1}{\eta_{\mathrm{AP}}}\sum_{j}c_{z,j}^{\mathrm{DL}},
\end{equation}
and the weighted energy efficiency (WEE) is
\begin{equation}\label{eq:wee}
\mathrm{WEE}(\mathbf{b})=\frac{B\,R_{w}(\mathbf{b})}{P_{\mathrm{static}}
+P_{\mathrm{FH}}+P_{\mathrm{BB}}+P_{\mathrm{TX}}}\ \ \text{[bit/Joule]},
\end{equation}
where $R_{w}=\sum_{k}(w^{\mathrm{UL}}\mathrm{SE}_{k}^{\mathrm{UL}}
+w^{\mathrm{DL}}\mathrm{SE}_{k}^{\mathrm{DL}})$ and $B$ is the bandwidth. Here, $w^{\mathrm{UL}}$ and $w^{\mathrm{DL}}$ are the positive weights that represent the importance of the uplink and downlink data rates, respectively.

\vspace{-1mm}

\section{Learned Split and Bit Allocation}\label{sec:learning}
\vspace{-1mm}

\subsection{Problem and Inputs}
\vspace{-1mm}

The design problem is
\vspace{-1mm}
\begin{equation}\label{eq:prob}
\max_{\mathbf{b}\in\mathcal{B}^{2LN}}\ \mathbb{E}\{\mathrm{WEE}(\mathbf{b})\}
\qquad\text{s.t.}\ \sum_{j=1}^{2LN}b_{j}\le B_{\mathrm{tot}}, \vspace{-1mm}
\end{equation}
which is a combinatorial problem over $(b_{\max}+1)^{2LN}$ points with a non-concave objective in which selection, resolution, and AP sleep are all coupled. Note that the constraint is an inequality since both $P_{\mathrm{FH}}$ and $P_{\mathrm{BB}}$ grow with the allocation and, beyond a certain budget, the energy-optimal policy leaves part of $B_{\mathrm{tot}}$ unused.
In this paper, we learn a mapping from the
large-scale statistics to $\mathbf{b}$, so that one forward pass replaces the
optimization. The inputs are, for every UE--AP pair, the edge feature
\begin{equation}\label{eq:edgefeat}
\bm{\chi}_{kl}=\big[\log_{10}(\beta_{kl}),\ d_{kl}/d_{\max},\
\sin(\theta_{kl}),\ \cos(\theta_{kl})\big]^{\Ttran}\in\mathbb{R}^{4},
\end{equation}
with $d_{kl}$ the two-dimensional AP--UE distance and $\theta_{kl}$ the azimuth angle, and, for
every (AP, beam) pair, the \emph{statistical} beam gains
\begin{equation}\label{eq:beamfeat}
\bm{\psi}_{ln}=\Big[\log_{10}\big(1+\textstyle\sum_{k}g_{kln}\big),\
\log_{10}\big(1+\max_{k}g_{kln}\big)\Big]^{\Ttran},
\end{equation}
where $g_{kln}=\mathbb{E}\{|[\mathbf{c}_{kl}]_{n}|^{2}\}
=[\mathbf{U}_{N}^{\Htran}\mathbf{R}_{kl}\mathbf{U}_{N}]_{nn}$ is the average
energy that DFT bin $n$ of AP $l$ collects from UE $k$. Both are large-scale,
location-dependent quantities, so no instantaneous channel state information is needed at the policy.

\subsection{Permutation-Equivariant Bipartite GNN}
The UEs and APs are the two vertex sets of a complete \emph{bipartite} graph
whose $KL$ edges carry $\bm{\chi}_{kl}$. Edge embeddings are initialized as
$\mathbf{e}_{kl}^{(0)}=\phi_{\mathrm{e}}(\bm{\chi}_{kl})\in\mathbb{R}^{h}$
and the AP/UE embeddings as their means over $k$ and $l$, respectively. They
are refined by $T$ rounds of residual message passing,
\begin{align}\label{eq:mp}
\mathbf{e}_{kl}^{(t)}&=\mathbf{e}_{kl}^{(t-1)}+\phi_{\mathrm{e}}^{(t)}
\big([\mathbf{e}_{kl}^{(t-1)},\mathbf{u}_{k}^{(t-1)},
\mathbf{a}_{l}^{(t-1)}]\big),\notag\\
\mathbf{a}_{l}^{(t)}&=\mathbf{a}_{l}^{(t-1)}+\phi_{\mathrm{a}}^{(t)}
\big([\overline{\mathbf{e}}_{\cdot l}^{(t)},\,
\max_{k}\mathbf{e}_{kl}^{(t)},\,\mathbf{a}_{l}^{(t-1)}]\big),\\
\mathbf{u}_{k}^{(t)}&=\mathbf{u}_{k}^{(t-1)}+\phi_{\mathrm{u}}^{(t)}
\big([\overline{\mathbf{e}}_{k\cdot}^{(t)},\,
\max_{l}\mathbf{e}_{kl}^{(t)},\,\mathbf{u}_{k}^{(t-1)}]\big),\notag
\end{align}
where $\overline{(\cdot)}$ denotes the mean over the omitted index, the
maxima are entrywise, and every $\phi$ is a two-layer ReLU multilayer
perceptron \emph{shared} by all UEs, APs and edges. Relabelling the UEs (or
APs) therefore permutes the outputs identically: the policy is
permutation-equivariant. The readout is
beam-resolved,
$(s_{ln}^{\mathrm{UL}},s_{ln}^{\mathrm{DL}})=\phi_{\mathrm{r}}
([\mathbf{a}_{l}^{(T)},\bm{\psi}_{ln}])$, giving $2LN$ scores, and a single
permutation-invariant \emph{null} score $s_{0}=\phi_{0}([\overline{\mathbf{a}}
^{(T)},\overline{\mathbf{u}}^{(T)}])+\log(2LN)$ is read out from the
mean-pooled embeddings.

\subsection{Budget-Feasible Discrete Head}\label{sec:head}
The $2LN$ beam scores are passed through a \emph{single} softmax
\emph{together with} $s_{0}$, so that
\begin{equation}\label{eq:softmax}
f_{j}=\frac{e^{s_{j}}}{\sum_{i}e^{s_{i}}+e^{s_{0}}},\qquad
\sum_{j}f_{j}=1-f_{0}<1 ,
\end{equation}
and the share $f_{0}$ of the null slot is not spent. Hence, the budget in \eqref{eq:prob} holds by construction, and the network learns both where to place the bits and how many bits to use. Since the utilization is determined by the same softmax, it is coupled to the beam scores. The offset $\log(2LN)$ balances the single null logit against the $2LN$ beam logits. The bits are then
\begin{align}\label{eq:head}
\widetilde{b}_{j}&=\big[B_{\mathrm{tot}}f_{j}\big]_{0}^{b_{\max}},\qquad
\overline{B}=\mathrm{round}\big(\textstyle\sum_{j}\widetilde{b}_{j}\big),
\notag\\
b_{j}&=\widetilde{b}_{j}+\big(\mathrm{LR}_{j}(\widetilde{\mathbf{b}},
\overline{B})-\widetilde{b}_{j}\big)_{\mathrm{detach}} ,
\end{align}
where $[\cdot]_{0}^{b_{\max}}$ is clipping and $\mathrm{LR}(\cdot)$ is the
largest-remainder projection, which floors every $\widetilde{b}_{j}$ and hands
the leftover to the largest fractional parts, so that exactly
$\overline{B}\le B_{\mathrm{tot}}$ integer bits are spent, and $(\cdot)_{\mathrm{detach}}$ is the stop-gradient operator, which passes its argument unchanged in the forward pass but treats it as a constant (i.e., with zero gradient) in the backward pass. The second line is
the straight-through estimator (STE) \cite{bengio2013estimating}: the forward
pass carries the exact integer allocation while the backward pass uses
$\partial b_{j}/\partial\widetilde{b}_{j}\triangleq1$. 

\subsection{Unsupervised Loss}\label{sec:loss}
No labels are used: for a batch $\mathcal{T}$ of UE drops the loss is
\begin{equation}\label{eq:loss}
\mathcal{L}(\bm{\Theta})=-\frac{1}{|\mathcal{T}|}\sum_{t\in\mathcal{T}}
\frac{1}{X}\sum_{x=1}^{X}\mathrm{WEE}\big(\mathbf{b}(\bm{\Theta};
\mathcal{I}_{t}),\mathbf{H}_{t}^{(x)}\big),
\end{equation}
i.e., the negated WEE with the rates Monte-Carlo averaged over $X$ Rayleigh
realizations per drop, an unbiased estimate of the ergodic objective, where $\mathcal{I}_{t}$ collects the input features $\{\bm{\chi}_{kl}\}$ and $\{\bm{\psi}_{ln}\}$ of drop $t$ and $\mathbf{H}_{t}^{(x)}$ is its $x$th channel realization. The
budget is enforced by the head and the power constraint by \eqref{eq:scale},
so training is an \emph{unconstrained} minimization solved with Adam. Two
quantities are piecewise constant in $\mathbf{b}$ and need a surrogate
gradient: $\alpha(b)$, whose forward pass uses the exact table entry while the
backward pass uses the \emph{central} difference
$\frac12(\rho(b+1)-\rho(b-1))$, and the counts $M^{\mathrm{UL}}$,
$M^{\mathrm{DL}}$, $A$, whose indicators are replaced by sigmoids centred on
the cut-off that the largest-remainder step actually used, again with an STE;
the ``AP is awake'' disjunction is taken probabilistically as
$1-\prod_{n}(1-\iota_{n})$ over the AP's $2N$ (smoothed) beam indicators, where $\iota_{n}\in(0,1)$ is the sigmoid surrogate of the indicator that the $n$th of these $2N$ beam slots is retained, i.e., receives at least one bit.

\subsection{Training}\label{sec:complexity}

We train with Adam (learning rate $2\cdot10^{-3}$, batch size $200$, at most
$500$ epochs with early stopping on the validation WEE, and a $10\times$
larger step size for the null head, which is a single scalar competing with
$2LN$ per-beam scores). One policy is trained per budget, from scratch, and
the best of three random restarts on the validation set is retained. 

\section{Evaluation, Benchmarks, and Results}\label{sec:results}

\subsection{Bussgang-Based Evaluation}
The AQNM is used for the \emph{design} only. In the evaluation, the
actual Lloyd--Max quantizers are applied to the beam-domain samples
generated by the data symbols, and the resulting nonlinearity is handled with
the Bussgang decomposition \cite{demir2021bussgang} taken with respect to the
symbol vector $\mathbf{s}$, while the combiner and the precoder are kept
exactly as designed with the AQNM statistics; the model mismatch is thus
reflected in the reported numbers. Denoting by $\widehat{s}_{k}$ the soft estimate of $s_{k}$ at the CPU (uplink)
or at the UE (downlink), the receiver relies only on the \emph{mean} effective
channel, i.e., we use the ergodic use-and-then-forget (UatF) bound \cite{cell-free-book}
\begin{align}\label{eq:uatf}
\mu_{k}&=\mathbb{E}\{\widehat{s}_{k}s_{k}^{*}\},\qquad
\mathrm{SINR}_{k}=\frac{|\mu_{k}|^{2}}
{\mathbb{E}\{|\widehat{s}_{k}-\mu_{k}s_{k}|^{2}\}},\notag\\
\mathrm{SE}_{k}&=\log_{2}(1+\mathrm{SINR}_{k}),
\end{align}
where the expectations are taken over \emph{both} the channel realizations and
the data symbols.

Although the CPU uses instantaneous CSI to construct the combiner and precoder, the reported rates use an ensemble UatF bound to obtain a robust achievable rate under the nonlinear quantizers without requiring the UEs/decoder to track the realization-dependent post-quantization effective channel.

\begin{figure*}[!t]
\centering
\includegraphics[width=\textwidth]{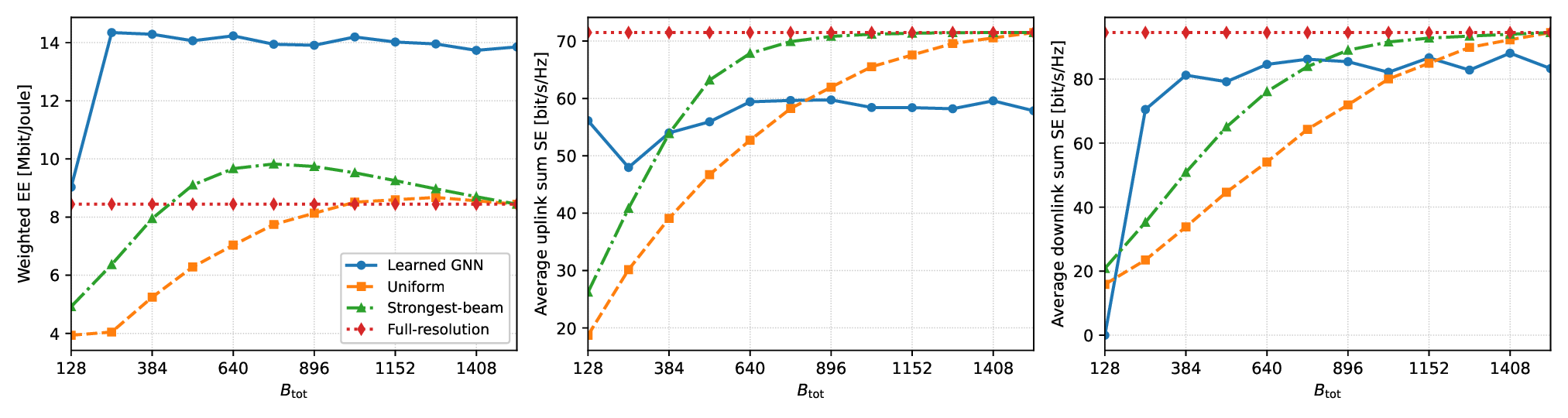}\vspace{-3mm}
\caption{(a) WEE (left), (b) average uplink sum SE (middle), and (c)
average downlink sum SE (right) versus the total fronthaul bit budget
$B_{\mathrm{tot}}$.}
\label{fig:res}
\vspace{-3mm}
\end{figure*}
\subsection{Benchmarks and Setup}
The learned policy is compared, under the same evaluation, with: (i)
\emph{uniform}, which spreads $B_{\mathrm{tot}}$ as evenly as possible over
all $2LN$ beams; (ii) \emph{strongest-beam}, which allocates the bits in
proportion to $\log_{2}(1+\sum_{k}|[\mathbf{c}_{kl}]_{n}|^{2})$, a
channel gain-aware but EE-agnostic heuristic; and (iii) \emph{full-resolution},
which forwards every beam at $b_{\max}$ bits, an infeasible reference that
isolates the quantization gap in the SE but does \emph{not} upper-bound the
WEE, since it inflates both $P_{\mathrm{FH}}$ and $P_{\mathrm{BB}}$.

We take $L=16$ APs on a regular grid over $1\times1$\,km$^{2}$, $N=8$
($2LN=256$ beam slots), $K=12$ UEs dropped uniformly at random. The large-scale fading follows a 3GPP Urban Microcell
non-line-of-sight (NLOS) model \cite[Table 7.4.1-1]{3gpp38901},
\begin{equation}
\beta_{kl}\,[\mathrm{dB}] = -32.4 - 31.9\log_{10}(\bar{d}_{kl})
                            - 20\log_{10}(f_c) + F_{kl},
\end{equation}
where $\bar{d}_{kl}$ is the three-dimensional AP--UE distance in meters, $f_c=3.5$ in GHz, and $F_{kl}\sim\mathcal{N}(0,8.2^2)$ is the shadow fading in dB. Spatial correlation matrices are generated using Gaussian local scattering model with $15^{\circ}$ angular spread for both azimuth and elevation \cite{cell-free-book}.
$B=100$\,MHz, $p=0.2$\,W, $P_{\max}=4$\,W, $b_{\max}=6$, $\lambda=5$,
$\xi_{\mathrm{FH}}=0.25\cdot10^{-8}$\,W/(bit/s), $\xi_{\mathrm{BB}}=2$\,W per
stream, $P_{\mathrm{AP}}=20$\,W, $P_{\mathrm{CPU}}=300$\,W,
$\eta_{\mathrm{sl}}=0.3$, $\eta_{\mathrm{AP}}=0.4$, $\eta_{\mathrm{UE}}=0.3$
and $w^{\mathrm{UL}}=w^{\mathrm{DL}}=1$. The GNN uses $h=96$ and $T=4$ and is
trained on $2000$ drops ($X=10$ realizations each), $200$ drops being used for
validation; a separate set of $20$ drops with $250$ realizations and $500$
symbols each is used for the Bussgang decomposition-based evaluation.

\subsection{Results}
In Fig.~\ref{fig:res}(a), we plot the WEE versus $B_{\mathrm{tot}}$. The learned policy outperforms all the benchmarks over the whole range, and the WEE gain is up to $3.4\times$, $2.1\times$, and $67\%$ compared to the uniform, strongest-beam, and full-resolution schemes, respectively. The WEE saturates at about $14$\,Mbit/Joule beyond $B_{\mathrm{tot}}\approx384$ since the policy stops spending the budget when the additional $P_{\mathrm{FH}}$ and $P_{\mathrm{BB}}$ outweigh the additional rate. On the other hand, the WEE of the heuristics, which always spend the whole budget, first increases and then decreases.

Figs.~\ref{fig:res}(b) and (c) show the uplink and downlink sum SEs. With less
than a third of the bits, the learned policy attains $76\%$ of the
full-resolution uplink sum SE and $90\%$ of its downlink sum SE, which is the intended trade-off of the WEE objective. At $B_{\mathrm{tot}}=128$, the learned policy allocates almost the entire budget to the uplink and effectively disables the downlink since, under such a limited budget, concentrating the bits on the uplink yields a higher WEE than splitting them between the two directions.

These gains can be explained by the structure of the learned allocation. Instead of spreading the budget, the policy concentrates it on the few DFT bins that a UE
actually illuminates, giving them a resolution high enough to keep
$\alpha(b)$ close to one, and switches the remaining beams off. Every beam it
switches off saves $\xi_{\mathrm{BB}}$ watts of baseband power and removes its
contribution from $R_{\mathrm{FH}}$, and once \emph{all} $2N$ beams of an AP
are off, the AP itself goes to sleep and its fixed power drops by
$(1-\eta_{\mathrm{sl}})P_{\mathrm{AP}}$. These three savings result from the same decision and, hence, the split, the resolution, and the AP activation should be jointly optimized. Furthermore, as $B_{\mathrm{tot}}$ increases, the softmax share of the null slot increases as well. Hence, the spent budget remains almost constant, whereas the benchmarks keep spending bits whose marginal rate does not compensate for their marginal power.

\section{Conclusion}\label{sec:conc}

In this paper, we have formulated the uplink split combining, the downlink split precoding, and the fronthaul bit allocation of a cell-free massive MIMO network as a single discrete decision. We have learned this decision end to end, without labels, by a permutation-equivariant bipartite GNN whose softmax head with a null slot satisfies the bit budget by construction and allows the network to spend only part of it. The simulation results, obtained with the actual Lloyd--Max quantizers and the Bussgang-based UatF bound, show that the learned policy improves the WEE by up to $3.4\times$ and $2.1\times$ compared to the uniform allocation and the channel gain-aware heuristic, respectively. We are planning to extend the framework to imperfect CSI and joint power control.

\bibliographystyle{IEEEtran}
\bibliography{IEEEabrv,refs}

\end{document}